\documentclass[aps,prb,twocolumn]{revtex4} 

\usepackage{graphicx}
\usepackage{dcolumn}
\usepackage{bm}

\newcommand{\comment}[1]{}

\begin{document}

\title{Quantum Statistical Mechanics in Classical Phase Space.
Test Results for Quantum Harmonic Oscillators}


\author{Phil Attard}
\affiliation{phil.attard1@gmail.com Sydney NSW, Australia}


\begin{abstract}
The von Neumann trace form of quantum statistical mechanics
is transformed to an integral over classical phase space.
Formally exact expressions for the resultant position-momentum
commutation function are given.
A loop expansion for wave function symmetrization is also given.
The method is tested for quantum harmonic oscillators.
For both the boson and fermion cases,
the grand potential and the average energy
obtained by numerical quadrature over classical phase space
are shown to agree with the known analytic results.
A mean field approximation is given
which is suitable for condensed matter,
and which allows the quantum statistical mechanics of interacting particles
to be obtained in classical phase space.
\end{abstract}


\maketitle

%
\section{Introduction}
\setcounter{equation}{0} \setcounter{subsubsection}{0}
\renewcommand{\theequation}{\arabic{section}.\arabic{equation}}
%

Many-particle systems pose serious
computational challenges for quantum mechanics.
These primarily arise from the difficulty in finding the energy
eigenfunctions and eigenvalues of the system,
and from the difficulty in enforcing the boson and fermion occupancy rules.
In addition to the technical barriers that inhibit
the accurate numerical description of many-particle systems,
it is the rapid increase in computational cost with system size
that can be prohibitive.\cite{Bloch08,Hernando13}
Common partial differential equation algorithms, for example,
scale exponentially with system size.\cite{Morton05}

Of course various and sophisticated attempts to ameliorate the difficulties
have been made,
such as an imaginary-time nonuniform mesh method,\cite{Hernando13}
pseudo-potential and mean-field methods,
\cite{Dalfovo95,Savenko13,Rogel13} 
density functional theory,\cite{Parr94,McMahon12}  
quantum Monte Carlo methods,
\cite{Pollet12,McMahon12,Mallory15,Ancilotto17} 
lattice Gaussian approach,\cite{Dimler09,Shimshovitz12,Machnes16}  
collocation method,\cite{Kosloff88} 
discrete variable representation method,\cite{Harris65} 
and variational Gaussian wave-packet methods,
\cite{Frantsuzov03,Frantsuzov04,Georgescu11a,Georgescu10,Georgescu11b}
as examples.
It is usually the case that various approximations
are introduced in these methods,
such as neglecting wave function symmetrization,
which can limit their individual reliability and range of application.

Despite in many cases the proven merits of these algorithms
and their demonstrated improvement over partial differential equation methods,
the size of quantum many body systems that can currently be treated
computationally
remains perhaps one or two orders of magnitude smaller
than classical many body systems.
For example,  Hernando and Van\'i\v cek
found the first 50 energy eigenvalues
and included symmetrization effects,
but the system contained
just five Lennard-Jones atoms in one dimension.\cite{Hernando13}
Larger system sizes have been achieved,
but at the cost of additional approximations
or  neglect of one or other quantum effect.
For example,
the variational treatment of up to 6500 Lennard-Jones atoms
by Georgescu and Mandelshtam was restricted to the ground state,
as well as neglecting wave function symmetrization.\cite{Georgescu11a}

Compared to quantum systems,
classical many-particle systems
scale  much more favorably with system size
and have considerably reduced computational demands.
This  suggests that an advantageous numerical approach for quantum systems
could be developed by performing
an expansion about the overlying classical system.

In previous work the author has presented
a transformation of the von Neumann trace form
for the quantum partition function and averages
to classical phase space.\cite{QSM,STD2}
The analysis invoked directly position and momentum states,
and is a somewhat simpler formulation
than the earlier method of Wigner,\cite{Wigner32}
and of Kirkwood.\cite{Kirkwood33}
The author's approach is not directly related
to these earlier approaches,
although they agree upon the first and second quantum
corrections to classical statistical mechanics.\cite{QSM,STD2}
The Wigner function has been used to explore
aspects of the quantum-classical relationship,
\cite{Curtright01,Bolivar04,Polkovnikov10}
as well as the related problem of the dissipative evolution
in open quantum systems,
\cite{Petruccione02,Zhang03,Bolivar12,Caldeira14,Cabrera15}
and the quantum-classical transition.
\cite{Habib02,Bhattacharya03,Zurek03,Everitt09,Jacobs14}
It has also been  used for quantum optics.
\cite{Groenewold46,Moyal49,Praxmeyer02,Barnett03,Gerry05,Zachos05,Dishlieva08}

In the author's approach,
quantum statistical mechanics is cast
as an integral over classical phase space
of the Maxwell-Boltzmann factor times
the product of two formally exact series expansions,
one that accounts for wave function symmetrization,
and the other for the non-commutativity of position and momentum operators.
\cite{QSM,STD2}
The leading term is precisely classical statistical mechanics.
This suggests that the approach might be both feasible and accurate
for condensed matter problems,
because for many terrestrial systems the quantum correction
amounts to no more than a fraction of a per cent.
Further, the approach represents a systematic approximation,
which is an advantage because
the error due to the truncation of the infinite series
can be quantified term by term.
Finally, since the method is cast
in terms of classical averages over phase space,
all of the techniques and algorithms that have been developed
over the years
for the computer treatment of classical many-particle systems
become immediately available for quantum systems.

The present work gives a simpler and more rigorous derivation of the
transformation to classical phase space than in the earlier work.
\cite{QSM,STD2}
It also derives new general expressions
for the statistical average of an operator,
and corrects several errors that appear in the earlier presentation.
In addition a new form for the so-called commutation function is given,
which appears to be more useful than previous high temperature expansions.

The present paper tests the classical phase space approach numerically
for the case of the quantum harmonic oscillator,
for which exact analytic results are well-known.
\cite{Messiah61,Merzbacher70,Pathria72}
As a concrete example,
the present results clarify and validate the new formulation
of quantum statistical mechanics,
which should lead to a better appreciation of its utility.
In addition, a mean field approximation
is derived based on the analytic results for the simple harmonic oscillator,
and this ought to be useful for general condensed matter,
interacting particle systems.


%
\section{The Partition Function and the Symmetrization Function}
\setcounter{equation}{0} \setcounter{subsubsection}{0}
%

\subsection{General Case of Interacting Particles}

Consider a system of $N$ interacting particles.
Let $| {\bf n} \rangle $
be an unsymmetrized, normalized  energy eigenfunction,
$\hat{\cal H}|{\bf n}\rangle = E_{\bf n} | {\bf n}\rangle$.
The position representation vector is
${\bf r} = \{{\bf r}_1,{\bf r}_2,\ldots,{\bf r}_N\}$,
with ${\bf r}_j = \{r_{jx},r_{jy},\ldots,r_{jd}\}$
in a space of $d$ dimensions.
Assume that the energy state can be
similarly decomposed,
${\bf n} = \{{\bf n}_1,{\bf n}_2,\ldots,{\bf n}_N\}$.
For simplicity spin is not here considered,
although its inclusion would not create  insurmountable difficulties.

Writing the unsymmetrized wave function also as
$\phi_{\bf n}({\bf r}) \equiv | {\bf n} \rangle $,
the symmetrized wave function formed from the orthonormal set of these is
\begin{eqnarray}
\phi_{\bf n}^\pm({\bf r})
& = &
\frac{1}{\sqrt{N! \chi^\pm_{\bf n}}}
\sum_{\hat{\mathrm P}} (\pm 1)^p \phi_{\hat{\mathrm P}{\bf n}}({\bf r})
\nonumber \\
\mbox{ or }
| {\bf n} \rangle^\pm
& = &
\frac{1}{\sqrt{N! \chi^\pm_{\bf n}}}
\sum_{\hat{\mathrm P}} (\pm 1)^p \, | \hat{\mathrm P}{\bf n} \rangle .
\end{eqnarray}
Here $p$ is the parity of the permutation $\hat{\mathrm P}$.
The upper sign is for bosons and the lower sign is for fermions.

The symmetrization factor $\chi^\pm_{\bf n}$
is characteristic of the state.
It is inversely proportional to the number of non-zero
distinct permutations of the wave function.
Specifically, normalization,
$\langle \phi_{\bf n}^\pm  | \phi_{\bf n}^\pm \rangle = 1$,
gives
\begin{equation}
\chi^\pm_{\bf n} =
\sum_{\hat{\mathrm P}} (\pm 1)^p
\langle {\bf n} |  \hat{\mathrm P}{\bf n} \rangle .
\end{equation}

The grand partition function is the sum over \emph{distinct} states,
\cite{QSM,STD2}
\begin{eqnarray}
\Xi^\pm
& = &
\sum_{N=0}^\infty z^N
\sum_{\bf n}\!'
e^{-\beta E_{\bf n}}
\nonumber \\ & = &
\sum_{N=0}^\infty \frac{z^N}{N!}
\sum_{\bf n} \chi_{\bf n}^\pm
e^{-\beta E_{\bf n}}
\nonumber \\ & = &
\sum_{N=0}^\infty \frac{z^N}{N!}
\sum_{\bf n}
\sum_{\hat{\mathrm P}} (\pm 1)^p
\langle {\bf n} | e^{-\beta \hat{\cal H}}
| {\hat{\mathrm P}{\bf n}}\rangle  .
\end{eqnarray}
Here the fugacity is $z \equiv e^{\beta \mu}$,
where $\mu$ is the chemical potential,
and $\beta = 1/k_\mathrm{B}T$  is called the inverse temperature,
with $k_\mathrm{B}$ being Boltzmann's constant and $T$ the temperature.
The sum over energy states in the final two equalities is unrestricted,
since the symmetrization factor accounts for double counting of the same state,
and for the cancelation of forbidden states.

The permutation operator that transposes particles $j$ and $k$
is $\hat{\mathrm P}_{jk}$.
Any permutation is a sequence of such pair transpositions,
the number of which gives its parity.
A connected sequence is called a loop,
(eg.\  $\hat{\mathrm P}_{jk}\hat{\mathrm P}_{kl}$ is a three particle loop).
Any permutation may be expressed as a product of loops.
Hence the symmetrization factor may be expanded as\cite{QSM,STD2}
\begin{eqnarray}
\chi^\pm_{\bf n}
& = &
\sum_{\hat{\mathrm P}} (\pm 1)^p
\langle \hat{\mathrm P} {\bf n} |  {\bf n} \rangle
\nonumber \\ & = &
1 \pm \sum_{j,k}^N\!'
\langle \hat{\mathrm P}_{jk}{\bf n} | {\bf n}  \rangle
+ \sum_{j,k,l}^N\!'
\langle \hat{\mathrm P}_{jk}\hat{\mathrm P}_{kl}{\bf n} | {\bf n}\rangle
\nonumber \\ & & \mbox{ }
+ \sum_{j,k,l,m}^N\!\!\!'\,
\langle \hat{\mathrm P}_{jk}\hat{\mathrm P}_{lm}{\bf n} | {\bf n}\rangle
\pm \ldots
\nonumber \\ & \equiv &
1 + \sum_{j,k}^N\!' \chi_{jk}^\pm({\bf n})
+ \sum_{j,k,l}^N\!' \chi_{jkl}^\pm({\bf n})
\nonumber \\ & & \mbox{ }
+ \sum_{j,k,l,m}^N\!\!\!'\,
\chi_{jk,lm}^\pm({\bf n})
+ \ldots
\end{eqnarray}
The prime indicates distinct permutations,
$\sum_{j,k}' \equiv \sum_{j<k}$ etc.
The comma in $\chi_{jk,lm}^\pm({\bf n})$ etc.\
denotes separate loops.

The first term of unity gives
the monomer term in the partition function,
\begin{equation}
\Xi_1 \equiv
\sum_{N=0}^\infty \frac{z^N}{N!}
\sum_{\bf n} e^{-\beta E_{\bf n}} .
\end{equation}
The ratio of the full to the monomer  partition function
is just the monomer average of the  symmetrization factor,
\begin{eqnarray}
\lefteqn{
\frac{\Xi^\pm}{\Xi_1}
} \nonumber \\
& = &
\left<\chi^\pm \right>_1
\nonumber \\ & = &
\left< 1
+ \sum_{jk}\!'  \chi_{jk}^{\pm}
+ \sum_{jkl}\!'   \chi_{jkl}^{\pm}
+ \sum_{jklm}\!'  \chi_{jk,lm}^{\pm}
+ \ldots \right>_1
\nonumber \\ & = &
1
+ \left< \frac{N!}{(N-2)!2}  \chi^{\pm(2)} \right>_1
+   \left< \frac{N!}{(N-3)!3}  \chi^{\pm(3)}\right>_1
\nonumber \\ &&  \mbox{ }
+ \frac{1}{2}
\left<  \frac{N!}{(N-2)!2} \chi^{\pm(2)} \right>_1^2
+ \ldots
\nonumber \\ & = &
\sum_{\{m_l\}}
\frac{1}{m_l!}
\prod_{l=2}^\infty
\left< \frac{N!}{(N-l)!l} \chi^{\pm(l)} \right>^{m_l}_1
\nonumber \\ & = &
\prod_{l=2}^\infty
\sum_{m_l=0}^\infty
\frac{1}{m_l!}
\left< \frac{N!}{(N-l)!l}  \chi^{\pm(l)} \right>^{m_l}_1
\nonumber \\ & = &
\prod_{l=2}^\infty
\exp
\left< \frac{N!}{(N-l)!l} \chi^{\pm(l)} \right>_1 .
\end{eqnarray}
The third and following equalities write the average of the product
of loops
as the product of the averages.
This is exact for ideal system,
because then the energy basis functions
factorize into the product of single particle functions.
More generally,
as is discussed in the phase space derivation below,
it is exact in the thermodynamic limit,
since, for example, $m$ uncorrelated loops scale as $V^{m}$,
whereas $m$ correlated loops scale as $V$,
where $V$ is the volume.\cite{QSM,STD2}
The combinatorial factor accounts for the number of unique loops
in each term;
$\chi^{\pm(l)}$ refers
to any one set of $l$ particles,
since all sets give the same average.
Explicitly,
the $l$-loop symmetrization factor here is
\begin{eqnarray} 
\chi^{\pm(l)}({\bf n}^{(l)})
& = &
(\pm 1)^{l-1}
\langle \{ {\bf n}_1, {\bf n}_2, \ldots , {\bf n}_l \} |
\{ {\bf n}_2, {\bf n}_3, \ldots , {\bf n}_1 \}  \rangle
\nonumber \\ & = &
(\pm 1)^{l-1}
\langle {\bf n}^{(l)}  | {\bf n}'^{(l)}  \rangle .
\end{eqnarray}
This depends only on the state of the $l$ particles involved.

The grand potential is $-k_\mathrm{B}T$ times
the logarithm of the partition function.
The monomer grand potential is given by
\begin{equation}
-\beta \Omega_1 = \ln \Xi_1 .
\end{equation}
The full grand potential is given by
\begin{eqnarray} \label{Eq:Omega(l)n}
-\beta \Omega^\pm
& = &
-\beta \Omega_1 + \ln \frac{\Xi^\pm}{\Xi_1}
\nonumber \\ & = &
-\beta \Omega_1
+ \sum_{l=2}^\infty
\left< \frac{N!}{(N-l)!l} \chi^{\pm(l)} \right>_1
\nonumber \\ & \equiv &
 -\beta \sum_{l=1}^\infty \Omega_l^\pm .
\end{eqnarray}
The final equality defines the $l$-mer grand potential.

\subsection{Ideal System of Non-Interacting Particles} \label{Sec:IdealXi}

Now this result is applied
to an ideal system comprising non-interacting particles.
In this case the total energy is
just the sum of that of the individual particle states,
\begin{equation}
E_{\bf n} = \sum_{j=1}^N  \varepsilon_{{\bf n}_j} .
\end{equation}
Also, the wave function factorizes,
$|{\bf n}\rangle = \prod_j  |{\bf n}_j\rangle$,
as does the loop symmetrization factor
\begin{equation} 
\chi^{\pm(l)}({\bf n}^{(l)})
=
(\pm 1)^{l-1} \langle  {\bf n}_l| {\bf n}_{1}  \rangle \prod_{j=1}^{l-1}
\langle  {\bf n}_j| {\bf n}_{j+1}  \rangle  .
\end{equation}

With these
the monomer partition function becomes
\begin{eqnarray}
\Xi_1
& = &
\sum_{N=0}^\infty \frac{z^N}{N!}
\prod_{j=1}^N
\sum_{{\bf n}_j}  e^{-\beta \varepsilon_{{\bf n}_j} }
\nonumber \\ & = &
\sum_{N=0}^\infty \frac{z^N}{N!}
\left\{ \sum_{{\bf n}_1}  e^{-\beta \varepsilon_{{\bf n}_1} } \right\}^N
\nonumber \\ & = &
\exp \left\{z \sum_{{\bf n}_1} e^{-\beta \varepsilon_{{\bf n}_1} }  \right\}.
\end{eqnarray}
Hence the monomer grand potential is given by
$ -\beta \Omega_1 =
z \sum_{{\bf n}_1} e^{-\beta \varepsilon_{{\bf n}_1} } $.

The dimer grand potential is given by
\begin{eqnarray}
\lefteqn{
-\beta \Omega^\pm_2 } \nonumber \\
& = &
\left< \frac{N!}{(N-2)!2} \chi^{\pm(2)} \right>_1
\nonumber \\ & = &
\frac{\pm 1}{\Xi_1}
\sum_{N=2}^\infty \frac{z^N}{N!}
\frac{N!}{(N-2)!2}
\sum_{{\bf n}_{1},{\bf n}_{2 }}
\left\{ \rule{0cm}{.4cm}
e^{-\beta \varepsilon_{{\bf n}_{1}}}
e^{-\beta \varepsilon_{{\bf n}_{2}}}
\right.\nonumber \\ && \mbox{ } \times \left. \rule{0cm}{.4cm}
\langle {\bf n}_{1}  | {\bf n}_{2} \rangle \,
\langle {\bf n}_{2}| {\bf n}_{1} \rangle
\right\}
\prod_{j=3}^{N} \sum_{{\bf n}_j} e^{-\beta \varepsilon_{{\bf n}_j}}
\nonumber \\ & = &
\frac{\pm 1}{\Xi_1}
\sum_{N=2}^\infty \frac{z^N}{(N-2)!2}
\sum_{{\bf n}_{1}} e^{-2\beta \varepsilon_{{\bf n}_{1}}}
\prod_{j=3}^{N}  \sum_{{\bf n}_j} e^{-\beta \varepsilon_{{\bf n}_j}}
\nonumber \\ & = &
\frac{\pm 1}{\Xi_1}
\sum_{N=2}^\infty \frac{z^N}{(N-2)!2}
\sum_{{\bf n}_{1}} e^{-2\beta \varepsilon_{{\bf n}_{1}}}
\left\{  \sum_{{\bf n}_j} e^{-\beta \varepsilon_{{\bf n}_j}}
\right\}^{N-2}
\nonumber \\ & = &
\frac{\pm \, z^2}{2}
\sum_{{\bf n}_{1}} e^{-2\beta \varepsilon_{{\bf n}_{1}}} .
\end{eqnarray}
Similarly, the trimer grand potential is given by
\begin{eqnarray}
\lefteqn{
-\beta \Omega^\pm_3
} \nonumber \\
& = &
\left< \frac{N!}{(N-3)!3} \chi^{\pm(3)} \right>_1
\nonumber \\ & = &
\frac{1}{\Xi_1}
\sum_{N=3}^\infty \frac{z^N}{(N-3)!3}
\sum_{{\bf n}_{1},{\bf n}_{2},{\bf n}_{3}}
e^{-\beta [ \varepsilon_{{\bf n}_{1}}
+ \varepsilon_{{\bf n}_{2}}
+ \varepsilon_{{\bf n}_{3}} ]}
\nonumber \\ && \mbox{ } \times
\langle {\bf n}_{1} | {\bf n}_{2} \rangle \,
\langle {\bf n}_{2} | {\bf n}_{3} \rangle\,
\langle {\bf n}_{3} | {\bf n}_{1} \rangle
\prod_{j=4}^{N} \sum_{{\bf n}_j} e^{-\beta \varepsilon_{{\bf n}_j}}
\nonumber \\ & = &
\frac{1}{\Xi_1}
\sum_{N=3}^\infty \frac{z^N}{(N-3)!3}
\sum_{{\bf n}_{1}} e^{-3\beta \varepsilon_{{\bf n}_{1}}}
\left\{
\sum_{{\bf n}_{1}} e^{-\beta \varepsilon_{{\bf n}_{1}}}
\right\}^{N-3}
\nonumber \\ & = &
\frac{z^3}{3}
\sum_{{\bf n}_{1}} e^{-3\beta \varepsilon_{{\bf n}_{1}}} .
\end{eqnarray}
Continuing in this fashion, the full grand potential for this system
of non-interacting single particle states is
\begin{eqnarray} \label{Eq:BOmega-ni}
-\beta \Omega^\pm & = &
-\beta \Omega_1
+ \sum_{l=2}^\infty
\frac{(\pm 1)^{l-1} z^l }{l}
\sum_{{\bf n}_{1}} e^{-l\beta \varepsilon_{{\bf n}_{1}}}
\nonumber \\ & = &
\sum_{l=1}^\infty
\frac{(\pm 1)^{l-1} z^l}{l}
\sum_{{\bf n}_{1}} e^{-l\beta \varepsilon_{{\bf n}_{1}}}.
\end{eqnarray}

For the case of the simple harmonic oscillator in $d$-dimensions,
the single particle energy is\cite{Pathria72,Messiah61,Merzbacher70}
\begin{equation}
\varepsilon_{{\bf n}_j}
=
\left[ \frac{d}{2} +  n_{jx} + n_{jy} + \ldots + n_{jd} \right]\hbar \omega,
\end{equation}
with $n_{j\alpha} = 0,1,2,\ldots$.
In this case the grand potential is
\begin{eqnarray} \label{Eq:BOmega-SHO}
-\beta \Omega^\pm & = &
\sum_{l=1}^\infty
\frac{ (\pm 1)^{l-1} z^l }{l}
e^{-dl\beta \hbar \omega /2}
\prod_{\alpha=x}^d
\sum_{n_\alpha=0}^\infty
e^{-l\beta \hbar \omega  n_\alpha}
\nonumber \\ & = &
\sum_{l=1}^\infty
\frac{(\pm 1)^{l-1} z^l }{l}
\left[
\frac{e^{-l\beta \hbar \omega /2}}{1-e^{-l\beta \hbar \omega}}
\right]^d.
\end{eqnarray}
Note that this diverges for $ z > e^{d\beta \hbar \omega /2}$.

\subsubsection{Text Book Derivation}

Although superficially different,
the general result derived above for non-interacting particles
may be shown to agree with the standard text book result,
and it allows for a novel interpretation
of the terms that occur in a series expansion of the latter.

Following \S 6.2 of Pathria,\cite{Pathria72}
single particle states labeled by
$ \varepsilon $
can be occupied by $N_\varepsilon=0,1,\ldots, N_\pm$ particles,
with $N_+ = \infty$ for bosons, and $N_-=1$ for fermions.
The grand partition function is the weighted sum over all possible occupancies
of each state,
\begin{eqnarray}
\Xi^\pm(z) & = &
\sum_{N_0=0}^{N_\pm} \sum_{N_1=0}^{N_\pm} \ldots \sum_{N_\infty=0}^{N_\pm}
z^{\sum_\varepsilon' N_\varepsilon}
e^{-\beta \sum_\varepsilon' \varepsilon N_\varepsilon }
\nonumber \\ & = &
\prod_{\varepsilon=0}^\infty\!'
\left\{
\sum_{N_\varepsilon=0}^{N_\pm}
z^{N_\varepsilon}
e^{-\beta \varepsilon N_\varepsilon } \right\}
\nonumber \\ & = &
\prod_{\varepsilon=0}^\infty \!'
\left[ 1 \mp z e^{-\beta \varepsilon } \right]^{\mp 1}
\nonumber \\ & = &
\prod_{n_x=0}^\infty \prod_{n_y=0}^\infty \ldots \prod_{n_d=0}^\infty
\left[ 1 \mp z e^{-\beta \varepsilon_{\bf n} } \right]^{\mp 1}.
\end{eqnarray}
The sums and products over energy is an abbreviated notation
that visits each state once,
as is explicit in the final equality.
This is Pathria's expression.
The grand potential is given by the logarithm of this,
a subsequent expansion of which yields
\begin{eqnarray} \label{Eq:BOmega-ni-conv}
-\beta \Omega^\pm
& = &
\mp 
\sum_{ {\bf n}_1 }
\ln \!\left[ 1 \mp z e^{-\beta \varepsilon_{{\bf n}_1} }  \right]
\nonumber \\ & = &
\pm \sum_{ {\bf n}_1 } \sum_{l=1}^\infty
\frac{(\pm z)^l }{l} e^{-\beta \varepsilon_{{\bf n}_1} l}
\nonumber \\ & = &
\sum_{l=1}^\infty \frac{(\pm 1)^{l-1} z^l }{l}
\sum_{ {\bf n}_1 }
 e^{-\beta \varepsilon_{{\bf n}_1} l}  .
\end{eqnarray}
This agrees with the above expression based on symmetrization loops,
Eq.~(\ref{Eq:BOmega-ni}).
Notice that between the two
there is a one-to-one correspondence for the terms
indexed by $l$.
The interpretation is that each such term arises
from the cyclic permutation of the group of particles involved.
For the case of the simple harmonic oscillator,
the  one-dimensional summation over symmetrization loops,
Eq.~(\ref{Eq:BOmega-SHO}),
is a somewhat simpler expression
than the conventional multi-dimensional sum over possible energy states,
the first equality in Eq.~(\ref{Eq:BOmega-ni-conv}).

Although the present approach and the text-book approach arrive at the same
result for ideal states,
the present approach proceeds from a rather different view-point,
namely that the symmetrization of the wave function is the fundamental axiom,
and that the occupancy of states (multiple for bosons, single for fermions)
is a quantity derived from it.
The present approach will be shown to be extremely useful
when the results are transformed to the continuum
that is classical phases space.
In the case of the continuum it is impossible
to define unambiguously discrete states and the occupancy thereof,
whereas the symmetrization of the wave function itself remains
a valid concept.

%
\section{Quantum Partition Function in Classical Phase Space} \label{Sec:qp}
\setcounter{equation}{0} \setcounter{subsubsection}{0}
%

This section transform the partition function
from a sum over energy states
to an integral over classical phase space
by invoking  directly position and momentum states.
This is not directly related to the formulation of Wigner,\cite{Wigner32}
although the commutation function that arises here
also appears in the earlier analysis.
Kirkwood added the first quantum
correction due to wave function symmetrization;\cite{Kirkwood33}
the infinite resummation of symmetrization loops
appears unique to the author's analysis.\cite{QSM,STD2}

The position representation for $N$ particles in $d$ dimensions is
${\bf q} = \{ {\bf q}_1, {\bf q}_2, \ldots, {\bf q}_N\}$,
with ${\bf q}_j = \{ q_{jx}, q_{jy}, \ldots, q_{jd}\}$.
The momentum eigenfunctions are\cite{Messiah61,STD2}
\begin{equation}
|{\bf p}\rangle
\equiv
\zeta_{\bf p}({\bf q})
=
\frac{e^{-\tilde{\bf p}\cdot{\bf q}/i \hbar}}{V^{N/2}} ,
\end{equation}
where the volume of the sub-system is  $V=L^d$.
The momentum eigenfunctions form a complete orthonormal set.
The momentum label ${\bf p}$ is a $dN$-dimensional integer,
and the corresponding continuum components are,
$\tilde {p}_{j\alpha} = {p}_{j\alpha}\Delta_p $.
The spacing between momentum states is
$\Delta_p = 2\pi \hbar /L$.\cite{Messiah61}

It is possible to take the continuum limit of these immediately.
It is also possible to introduce  position eigenfunctions
that are Dirac-$\delta$ functions,
which form a complete orthonormal set.
In both cases the final phase space expression
is the same as that which results from the present analysis.

In order to re-sum the symmetrization  loops,
it is convenient to work in a grand canonical system,
in which the sub-system can exchange number and energy with a reservoir.
Following entanglement and collapse,
the grand canonical partition function has trace form.\cite{QSM,STD2}
Hence it can be written as a sum over the momentum states,
\begin{eqnarray}
\Xi^\pm
& = &
\sum_{N=0}^\infty \frac{z^N}{N!} \sum_{\hat {\mathrm P}} (\pm 1)^p \sum_{\bf p}
\left\langle \hat  {\mathrm P} {\bf p} \left|
e^{-\beta \hat {\cal H} }
\right| {\bf p} \right\rangle
\nonumber \\ & = &
\sum_{N=0}^\infty \frac{z^N}{N!} \Delta_p^{-dN} L^{-dN}
\sum_{\hat P} (\pm 1)^p
\nonumber \\ && \mbox{ } \times
\int \mathrm{d}{\bf q} \, \mathrm{d}{\bf p}\;
e^{(\hat{\mathrm P} \tilde{\bf p})\cdot{\bf q}/i \hbar}
e^{-\beta \hat {\cal H} }
e^{-\tilde{\bf p}\cdot{\bf q}/i \hbar}
\nonumber \\ & \equiv &
\sum_{N=0}^\infty \frac{z^N}{h^{dN}N!}
\int\!\! \mathrm{d}{\bf \Gamma}
 e^{-\beta {\cal H}({\bf \Gamma})}\,
 W_p({\bf \Gamma})\,
 \eta_q^\pm({\bf \Gamma}).
\end{eqnarray}
Here and below,
${\bf \Gamma} = \{ {\bf p},{\bf q}\}$
denotes a point in classical phase space,
and the tilde has been dropped.

Following Wigner,\cite{Wigner32}
the commutation function $W$
is defined via the action of the Hamiltonian operator
on the momentum eigenfunctions,\cite{QSM,STD2}
\begin{equation} \label{Eq:Wdef}
e^{-\beta \hat{{\cal H}} } e^{-{\bf p}\cdot{\bf q}/i \hbar}
\equiv
e^{-{\bf p}\cdot{\bf q}/i \hbar}
e^{-\beta {\cal H}({\bf p},{\bf q}) } W_p({\bf p},{\bf q}).
\end{equation}

As for the energy states,
the symmetrization function
may be expanded as\cite{QSM,STD2}
\begin{eqnarray}
\eta_q^\pm({\bf \Gamma})
& \equiv &
\sum_{\hat{\mathrm P}} (\pm 1)^p
e^{-{\bf p}\cdot{\bf q}/i \hbar}
e^{(\hat{\mathrm P} {\bf p})\cdot{\bf q}/i \hbar}
\nonumber \\ & = &
1 \pm \sum_{j,k}^N\!'
e^{-{\bf p}\cdot{\bf q}/i \hbar}
e^{(\hat{\mathrm P}_{jk} {\bf p})\cdot{\bf q}/i \hbar}
\nonumber \\ & & \mbox{ }
+ \sum_{j,k,l}^N\!'
e^{-{\bf p}\cdot{\bf q}/i \hbar}
e^{ (\hat{\mathrm P}_{jk}\hat{\mathrm P}_{kl} {\bf p})\cdot{\bf q}/i \hbar}
\pm \ldots
\nonumber \\ & = &
1 \pm \sum_{j,k}^N\!'
e^{-{\bf p}_j\cdot {\bf q}_{jk}/i \hbar}
e^{-{\bf p}_k\cdot {\bf q}_{kj}/i \hbar}
\nonumber \\ & & \mbox{ }
+ \sum_{j,k,l}^N\!'
e^{-{\bf p}_j\cdot {\bf q}_{jk}/i \hbar}
e^{-{\bf p}_k\cdot {\bf q}_{kl} /i \hbar}
e^{-{\bf p}_l\cdot {\bf q}_{lj} /i \hbar}
 \pm \ldots
\nonumber \\ & \equiv &
1 + \sum_{j,k}^N\!' \eta_{jk}^\pm({\bf \Gamma})
+ \sum_{j,k,l}^N\!' \eta_{jkl}^\pm({\bf \Gamma})
+ \ldots
\end{eqnarray}
Because these terms are highly oscillatory,
they average to zero unless consecutive particles
in a loop are close together
in position or momentum space.

\subsubsection{Grand Potential}

The series for the grand potential based on the loop expansion
obtained above via energy states, Eq.~(\ref{Eq:Omega(l)n})
carries over essentially unchanged.
The monomer grand potential is given by $-\beta \Omega_1 = \ln \Xi_1$,
with the monomer grand partition function being
\begin{equation}
\Xi_1
= \sum_{N=0}^\infty \frac{z^N}{h^{dN}N!}
\int \mathrm{d}{\bf \Gamma}\;
 e^{-\beta {\cal H}({\bf \Gamma})}\,
 W_p({\bf \Gamma}) .
\end{equation}
The loop grand potential is a monomer average in phase space,
\begin{eqnarray} \label{Eq:Oml}
\lefteqn{
-\beta \Omega_l^\pm
}  \\
& = &
\left< \frac{N!}{(N-l)!l} \eta^{\pm(l)} \right>_1 ,
\;\; l\ge 2
\nonumber \\ & = &
\frac{1}{\Xi_1}
\sum_{N=l}^\infty \frac{z^Nh^{-dN}}{(N-l)!l}
\int \mathrm{d}{\bf \Gamma}\;
 e^{-\beta {\cal H}({\bf \Gamma})}\,
 W_p({\bf \Gamma})\,
 \eta_q^{\pm(l)}({\bf \Gamma}^{l}).\nonumber
\end{eqnarray}
In obtaining this result the monomer average of the product of distinct loops
has been written as the product of the average of the individual loops,
which is valid because the individual loops must be compact
to avoid cancelation by rapid  oscillation.

The $l$-loop symmetrization function is
\begin{equation}
\eta_q^{\pm(l)}({\bf \Gamma}^l)
=
(\pm 1)^{l-1}
e^{ -{\bf q}_{l 1} \cdot {\bf p}_l /i\hbar }
\prod_{j=1}^{l-1}
e^{ -{\bf q}_{j,j+1}  \cdot {\bf p}_j /i\hbar } .
\end{equation}

The monomer term, $\eta^{\pm(1)} = 1$,
is obviously the classical one,
and it is the dominant term when $\rho \Lambda^d \ll 1$,
which is the low density $\rho$,
small thermal wave length $\Lambda = [2\pi \hbar^2 \beta/m]^{1/2}$
(or high temperature) limit.\cite{Kirkwood33,QSM,STD2}


The original von Neumann trace for the partition function is real.
The present transformation to classical phase space
introduces an asymmetry between position and momentum,
which induces an imaginary component.
But since this is odd in momentum,
it integrates to zero.
One can symmetrize the expression for the grand potential
with respect to position and momentum.
Since $ W_p^* = W_q$ and $\eta_q^{\pm*}=\eta_p^\pm$,
this is equivalent to making the replacement
$ W_p \eta_q^\pm  \Rightarrow \mbox{Re}(W_p\eta_q^\pm)$.
It is not essential to do this
because the imaginary parts of the integrand are odd in momentum
and so they integrate to zero.
(See Ref.~\onlinecite{Attard18} for a more detailed treatment of this point.)

%
\section{Expressions for the Commutation Function} \label{Sec:Ww}
\setcounter{equation}{0} \setcounter{subsubsection}{0}
%

\subsection{Expansion for Large $W$}

The commutation function $W_p$ was defined in Eq.~(\ref{Eq:Wdef}),
or equivalently
$ e^{{\bf p}\cdot{\bf q}/i\hbar}
e^{-\beta \hat{{\cal H}} }
e^{-{\bf p}\cdot{\bf q}/i\hbar}
=
e^{-\beta {\cal H}({\bf p},{\bf q}) } W_p({\bf p},{\bf q})$.
The subscript $p$ will now be dropped.
The Hamiltonian operator is
$\hat{\cal H} = {\cal H}( \hat {\bf p}, \hat {\bf q})
= U(\hat{\bf q}) + \hat p^2/2m$,
and the momentum operator is $ \hat {\bf p} = -i\hbar \nabla_{\bf q}$.

Following 
Kirkwood,\cite{Kirkwood33}
differentiation with respect to inverse temperature gives\cite{QSM,STD2}
\begin{eqnarray}
\frac{\partial W}{\partial \beta}
& = &
\frac{- \beta\hbar^2}{2m}  (\nabla^2 U) W
- \frac{\beta\hbar^2}{m} (\nabla U) \cdot (\nabla W)
\nonumber \\ & & \mbox{ }
 + \frac{\beta^2 \hbar^2}{2m} (\nabla U) \cdot (\nabla U) W
+ \frac{\hbar^2}{2m} \nabla^2 W
\nonumber \\ & & \mbox{ }
+ \frac{i\hbar}{m} {\bf p} \cdot  (\nabla W)
- \frac{i\hbar\beta}{m}  {\bf p} \cdot (\nabla U) W .
\end{eqnarray}
Here and below $\nabla \equiv \nabla_{\bf q}$.

Expand the commutation function in powers of inverse temperature,
\begin{equation}
W
=
\sum_{n=0}^\infty W_n \beta^n .
\end{equation}
One has $W_0=1$, which is the classical limit,
and $W_1=0$, since there are no terms of order $\beta^0$
on the right hand side of the temperature derivative.
Terms of order $\beta^1$ yields
\begin{equation}
W_2
=
\frac{-\hbar^2}{4m} \nabla^2 U
- \frac{i\hbar}{2m}  {\bf p} \cdot \nabla U,
\end{equation}
and those  of order $\beta^2$ give
\begin{eqnarray}
W_3  & = &
\frac{\hbar^2}{6m} \nabla U \cdot \nabla U
- \frac{\hbar^4}{24m^2} \nabla^2 \nabla^2 U
- \frac{i\hbar^3}{6m^2} {\bf p} \cdot \nabla \nabla^2 U
\nonumber \\ & & \mbox{ }
+ \frac{\hbar^2}{6m^2} {\bf p}{\bf p} : \nabla \nabla U.
\end{eqnarray}

Equating terms of order $\beta^n$ yields the recursion relation
\begin{eqnarray}
\lefteqn{
W_{n+1}
} \nonumber \\
& = &
\frac{- \hbar^2}{2(n+1)m}  (\nabla^2 U) W_{n-1}
- \frac{\hbar^2}{(n+1)m} \nabla U \cdot \nabla W_{n-1}
\nonumber \\ & & \mbox{ }
 + \frac{\hbar^2}{2(n+1)m} (\nabla U\cdot \nabla U) W_{n-2}
\nonumber \\ & & \mbox{ }
+ \frac{\hbar^2}{2(n+1)m} \nabla^2 W_{n}
+ \frac{i\hbar}{(n+1)m} {\bf p} \cdot  \nabla W_{n}
\nonumber \\ & & \mbox{ }
- \frac{i\hbar}{(n+1)m}  {\bf p} \cdot (\nabla U) W_{n-1} .
\end{eqnarray}

\subsection{Expansion for Small $w$}

Also defined has been a `small $w$' commutation function,
$W({\bf \Gamma}) = e^{w({\bf \Gamma})}$,
in the hope that its expansion might have better convergence properties.
\cite{QSM,STD2}
An expansion of this in powers of Planck's constant,
\begin{equation} \label{Eq:w=wnhn}
w \equiv \sum_{n=1}^\infty w_n \hbar^n ,
\end{equation}
and temperature differentiation
leads to the recursion relation
\begin{eqnarray}
\frac{\partial w_n}{\partial \beta }
& = &
\frac{i}{m} {\bf p} \cdot \nabla w_{n-1}
 + \frac{1}{2m}
 \sum_{j=1}^{n-2}  \nabla w_{n-2-j}  \cdot \nabla w_j
\nonumber \\ && \mbox{ }
- \frac{\beta}{m} \nabla w_{n-2}  \cdot \nabla U
 + \frac{1}{2m} \nabla^2 w_{n-2} .
\end{eqnarray}
The first several terms may be obtained explicitly.
One has for $n=0$, $w_0 =0$,
for $n=1$,
\begin{equation} \label{Eq:w1}
 w_1
=
\frac{-i\beta^2}{2m} {\bf p} \cdot \nabla  U ,
\end{equation}
for $n=2$,
\begin{equation} \label{Eq:w2}
w_2
=
\frac{\beta^3}{6m^2}
{\bf p} {\bf p} : \nabla \nabla  U
 + \frac{1}{2m}
\left\{ \rule{0cm}{0.4cm}
\frac{ \beta^3}{3}  \nabla  U \cdot \nabla  U
-\frac{ \beta^2}{2}  \nabla^2 U
\rule{0cm}{0.4cm}\right\} ,
\end{equation}
for $n=3$,
\begin{eqnarray}
w_3
& = &
\frac{i\beta^4}{24m^3}
{\bf p} {\bf p} {\bf p} \vdots \nabla\nabla \nabla  U
+ \frac{5i\beta^4}{24m^2}  {\bf p}  (\nabla  U) : \nabla \nabla  U
\nonumber \\ && \mbox{ }
-\frac{i\beta^3}{6m^2} {\bf p} \cdot \nabla \nabla^2 U ,
\end{eqnarray}
and for $n=4$,
\begin{eqnarray}
w_4
& = &
\frac{-\beta^5}{5!m^4} ({\bf p}\cdot\nabla)^4 U
+ \frac{\beta^4}{16m^3}  ({\bf p} \cdot \nabla)^2  \nabla^2 U
\nonumber \\ && \mbox{ }
- \frac{\beta^5 }{15m^3}
({\bf p} \cdot \nabla \nabla U)  \cdot ({\bf p} \cdot \nabla \nabla U)
\nonumber \\ && \mbox{ }
- \frac{3\beta^5 }{40m^3}
{\bf p} {\bf p} \nabla U \vdots \nabla\nabla\nabla U
- \frac{\beta^3}{24m^2} \nabla^2 \nabla^2 U
\nonumber \\ &  & \mbox{ }
+ \frac{5\beta^4}{48m^2} \nabla U \cdot \nabla \nabla^2 U
+ \frac{\beta^4}{24m^2} \nabla \nabla U : \nabla \nabla U
\nonumber \\ && \mbox{ }
- \frac{\beta^5}{15m^2} \nabla \nabla U : \nabla U \nabla U .
\end{eqnarray}
This result for $w_4$ corrects Eq.~(7.112) of Ref.~\onlinecite{STD2}.
Unfortunately published numerical results are vitiated by that error.
\cite{STD2,Attard17}

\subsubsection{Simple Harmonic Oscillator}

Using $\hbar \omega$ as the unit of energy,
and dimensionless momentum and position operators,
$\hat P = \hat p /\sqrt{ m \hbar \omega }$
and
$  \hat Q = \sqrt{m\omega/\hbar} \hat q $,
the Hamiltonian operator for the simple harmonic oscillator may be written
\cite{Messiah61}
\begin{equation}
\hat{\cal H}
= \frac{1}{2} \left\{ \hat P^2 + \hat Q^2 \right\} .
\end{equation}
In these dimensionless units, $\hbar = m = 1$.

For one particle in $d$ dimensions, $\nabla U = {\bf Q}$ and  $\nabla^2 U = d$.
Hence with $P^2\equiv {\bf P}\cdot {\bf P}$,
$Q^2 \equiv {\bf Q}\cdot {\bf Q}$, and  $R \equiv {\bf P}\cdot {\bf Q}$,
the `big $W$' recursion relation for the temperature
expansion coefficients of the simple harmonic oscillator is
\begin{eqnarray}
W_{n+1}
& = &
\frac{-  d}{2(n+1)} W_{n-1}
- \frac{1}{n+1} {\bf Q} \cdot \nabla W_{n-1}
\nonumber \\ & & \mbox{ }
 + \frac{ Q^2}{2(n+1)} W_{n-2}
+ \frac{1}{2(n+1)} \nabla^2 W_{n}
\nonumber \\ & & \mbox{ }
+ \frac{i}{n+1} {\bf P} \cdot  \nabla W_{n}
- \frac{iR}{n+1}  W_{n-1} .
\end{eqnarray}
The first several expansion coefficients are explicitly
$W_0  = 1$,  $W_1=0$, and
\begin{eqnarray}
W_2 & = &
\frac{-d}{4} - \frac{i}{2}R
 \\
W_3 & = &
\frac{1}{6} P^2 + \frac{1}{6}  Q^2
\nonumber \\
W_4 & = &
\frac{3d^2+4d}{96}
+ \frac{3d + 5 }{24}  i R
- \frac{1}{8} R^2
\nonumber \\
W_5 & = &
- \frac{5d + 8 }{120} P^2
- \frac{5d + 8 }{120} Q^2
- \frac{1}{12} iRQ^2
- \frac{1}{12} iRP^2 .\nonumber
\end{eqnarray}
Each of these is symmetric in ${\bf P}$ and ${\bf Q}$,
which is a useful check.

The first several `small $w$' coefficients in its expansion
in powers of Planck's constant
for the simple harmonic oscillator
for one particle in $d$ dimensions are explicitly
(with $\beta$ dimensionless)
\begin{eqnarray}
 w_1 & = &
\frac{-i \beta^2  }{2} R
\nonumber \\
w_2 & = &
\frac{ \beta^3 }{6 }   P^2
+ \frac{\beta^3  }{6} Q^2
- \frac{ d\beta^2  }{4}
\nonumber \\
 w_3 & = &
\frac{5i\beta^4 }{24} R
\nonumber \\
w_4 & = &
\frac{- \beta^5 }{15 } P^2
+ \frac{d\beta^4}{24}
- \frac{\beta^5}{15} Q^2.
\end{eqnarray}


\subsection{Series in Energy States}\label{Sec:Wn}

Now the commutation function will be expressed in energy eigenfunctions,
which will give a formally exact phase space representation.

The energy eigenfunctions and eigenvalues are
$\hat{\cal H} |{\bf n}\rangle = E_n  |{\bf n}\rangle$.
At this stage there is no need to be specific about the dimensionality
or the number of particles.
Formally the commutation function can be written as
\begin{eqnarray}
e^{-\beta{\cal H}({\bf p},{\bf q})} W_p({\bf p},{\bf q})
& = &
e ^{{\bf p}\cdot{\bf q}/i\hbar}
e^{-\beta\hat{\cal H}} e ^{-{\bf p}\cdot{\bf q}/i\hbar}
\\ & = &
e ^{{\bf p}\cdot{\bf q}/i\hbar}
L^{Nd/2} \sum_{\bf n}
e^{-\beta\hat{\cal H}} |{\bf n}\rangle \,
\langle{\bf n} |{\bf p}\rangle
\nonumber\\ & = &
e ^{{\bf p}\cdot{\bf q}/i\hbar}
L^{Nd/2} \sum_{\bf n }
e^{-\beta E_{\bf n}} \,
\langle {\bf n} |{\bf p}\rangle
\, \phi_{\bf n} ({\bf q}) .\nonumber
\end{eqnarray}
This expression is general and is not restricted
to ideal systems or to the simple harmonic oscillator.
In the summand appear in essence the energy eigenfunctions
and their Fourier transform,
\begin{eqnarray}
\langle {\bf n} |{\bf p}\rangle
& = &
L^{-Nd/2}
\int \mathrm{d}{\bf q} \; e^{-{\bf p}\cdot {\bf q}/i\hbar}
\phi_{\bf n}({\bf q})
\nonumber \\ & \equiv &
L^{-Nd/2} \check \phi_{\bf n}({\bf p}),
\end{eqnarray}
With this
the weighted commutation function may be written
\begin{equation}
e^{-\beta{\cal H}({\bf p},{\bf q})}
W_p({\bf p},{\bf q})
=
e^{{\bf p}\cdot {\bf q}/i\hbar}
\sum_{\bf n }e^{-\beta E_{\bf n}}
 \check \phi_{\bf n}({\bf p})
\phi_{\bf n}({\bf q}) .
\end{equation}
The imaginary part of this is odd in ${\bf p}$.
This result is formally exact.


For an interacting system, one can imagine evaluating this expression for the
commutation function
by approximating the energy eigenvalues and eigenfunctions,
and then using it in addition to 
the symmetrization function
to weight the points of classical phase space.
(Alternatively, see \S \ref{Sec:HarmInt}.)

\subsubsection{Simple Harmonic Oscillator Commutation Function}
\label{Sec:Wsho}

This form for the commutation function can be given explicitly
for the simple harmonic oscillator.
In dimensionless units,
the energy eigenvalues are $E_{\bf n} = \sum_{j,\alpha} [n_{j\alpha}+1/2]$,
and the energy eigenfunctions are the Hermite functions,\cite{WikiQHO}
\begin{equation}
\phi_{\bf n}({\bf Q}) \equiv
\prod_{j,\alpha}
\frac{1}{\sqrt{2^{n_{j\alpha}} {n_{j\alpha}}! \sqrt{\pi}}}
e^{-Q_{j\alpha}^2/2}
\mathrm{H}_{n_{j\alpha}}(Q_{j\alpha}) ,
\end{equation}
where $\mathrm{H}_n(Q)$ is the Hermite polynomial of degree $n$.
The  Hermite function is essentially its own Fourier transform,
\begin{equation}
 \check \phi_{\bf n}({\bf P}) =
\prod_{j,\alpha}
\frac{i^{n_{j\alpha}} \sqrt{2\pi}
}{
\sqrt{2^{n_{j\alpha}} {n_{j\alpha}}! \sqrt{\pi}} }
e^{-P_{j\alpha}^2/2}
\mathrm{H}_{n_{j\alpha}}(P_{j\alpha}) .
\end{equation}

\begin{figure}[t!]
\centerline{
\resizebox{8cm}{!}{ \includegraphics*{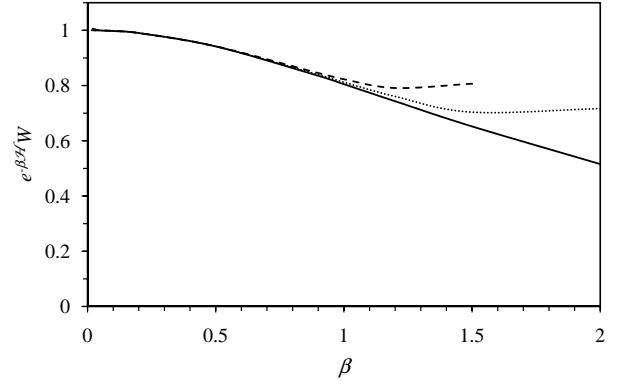} } }
\caption{\label{Fig:eW(0)-B}
The weighted commutation function,
$e^{-\beta {\cal H}(P,Q)} W(P,Q)$,
of the simple harmonic oscillator
at $P=Q=0$ as a function of inverse temperature $\beta$.
The solid curve is the exact result,
the dashed curve is the high temperature expansion
using all coefficients up to $W_5$,
and the dotted curve is the high temperature expansion $W=e^w$
using all coefficients up to $w_4$.
Note that ${\cal H}(0,0) = 0$,
and that $W(0,0)$ is real.
}
\end{figure}

Figure~\ref{Fig:eW(0)-B} shows the simple harmonic oscillator
commutation function for one particle in one dimension at $P=Q=0$.
One can see that the phase space weight is increasingly reduced
from the classical Maxwell-Boltzmann weight
as the temperature is decreased.
At higher temperatures, $\beta \alt 1$,
there is good agreement between the exact series form
and the six (up to $W_5$) and five (up to $w_4$)
term high temperature expansions
for $W$ and $w$, respectively.
The exact series used up to $n=54$ terms for high  temperatures,
$\beta \rightarrow 0$.
As the temperature is reduced the number of necessary terms declines:
at $\beta = 1$ the tenth term was ${\cal O}(10^{-10})$,
and at $\beta=2$ the fifth term was ${\cal O}(10^{-10})$.
The commutation function can differ significantly from unity at lower
temperatures;
at $\beta =2$ and $P=Q=0$,
it reduces the phase space weight
by about 50\% from its classical value.


\begin{figure}[t!]
\centerline{
\resizebox{8cm}{!}{ \includegraphics*{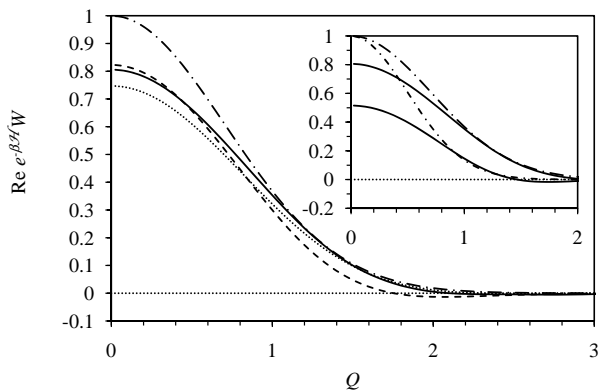} } }
\caption{\label{Fig:eW-Q}
Real part of the weighted commutation function,
$\mbox{Re}\, e^{-\beta {\cal H}(P,Q)} W(P,Q)$,
along the line $P=Q$ at $\beta =1.0$.
The dash-dotted curve is 
$e^{-\beta {\cal H}(P,Q)}$,
the solid curve is the exact result using $n=4$ terms,
the dashed curve is the high temperature expansion
using all coefficients up to $W_5$,
the dotted curve is is the high temperature expansion
using all coefficients up to $w_4$.
The horizontal dotted line is a guide to the eye.
{\bf Inset.} Results for $\beta=1$ (above) and $\beta=2$ (below).
}
\end{figure}

Figure~\ref{Fig:eW-Q} shows
the real part of the weighted commutation function,
$\mbox{Re}\, e^{-\beta {\cal H}(P,Q)} W(P,Q)$,
as a function of phase space along the line $P=Q$
at a fixed temperature $\beta = 1$.
The inset compares results for  $\beta = 1$ with  $\beta = 2$.
The commutation function generally decreases the phase space weight
from that given by the classical Maxwell-Boltzmann factor alone.
The effect is most significant in the region of the potential minimum,
in this case $P=Q \alt 1$,
or, equivalently,  ${\cal H} \alt 1$.
It can be seen in the main figure that the high temperature
expansions remain relatively accurate at $\beta = 1$.
Fewer terms need be retained in the exact series
as the temperature is decreased;
at $\beta = 2$ results with $n=4$ terms were indistinguishable
from those with $n=5$--$10$.

Some oscillatory behavior was observed at larger energies,
which depended upon the number of terms retained in the series.
One would like a probability density to be non-negative,
but there seems to be no fundamental requirement
that the transformation of quantum statistical mechanics
to classical phase space should yield an actual probability density.
In any case, the numerical problems
(ie.\ the sensitivity to the number of retained terms)
with the exact series for the commutation function at large energies
and high temperatures are likely moot
because the Maxwell-Boltzmann factor makes the total phase space weight
negligible in this regime.

\begin{figure}[t!]
\centerline{
\resizebox{8cm}{!}{ \includegraphics*{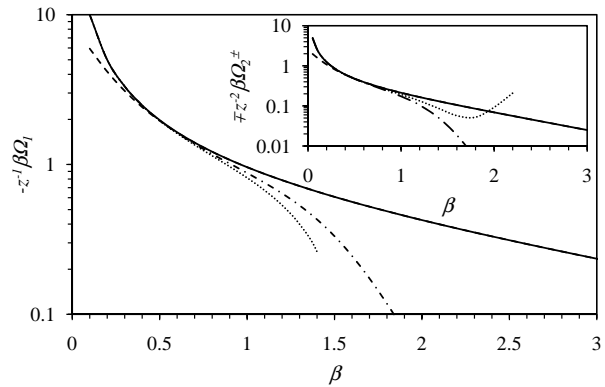} } }
\caption{\label{Fig:OmB}
The monomer grand potential
for ideal oscillators 
as a function of inverse temperature.
The solid curve is the exact analytic result,
Eq.~(\ref{Eq:BOmega-SHO}).
The remaining curves result from numerical quadrature over phase space,
Eq.~(\ref{Eq:Oml}),
using for the commutation function
the series over energy states up to $n=4$
(dashed curve, coincident with the analytic data at low temperatures),
the high temperature expansions up to $W_5$
(dotted curve)
and up to $w_4$
(dash-dotted curve),
both coincident  with the analytic data at high temperatures.
{\bf Inset.}
Corresponding results for the dimer grand potential.
}
\end{figure}

Figure~\ref{Fig:OmB} compares the grand potential
obtained as a phase space integral, 
Eq.~(\ref{Eq:Oml}),
with  the exact analytic result, Eq.~(\ref{Eq:BOmega-SHO}).
The imaginary parts of the phase space expressions integrate to zero.
As expected, the energy series expressions with a small number of terms
works well at low temperatures,
and the two high temperature expansions work well in the opposite regime.
The purpose of the figure is to show that it is quite feasible
to obtain computationally
the quantum grand potential from an integral over classical phase space.

Because the grand potential diverges in the classical limit,
$\beta \rightarrow 0$,
it can be more challenging to obtain accurate numerical results
in the high temperature regime.
At $\beta=0.2$, for the monomer grand potential $-z^{-1} \beta \Omega_1$,
the exact analytic result is 4.99,
using the energy series with $n=4$ terms it is 3.16,
and with $n=8$ terms it is 4.17.
Again at $\beta=0.2$, for the dimer grand potential
$\mp z^{-2} \beta \Omega^\pm_2$,
the exact analytic result is 1.24,
using the energy series with $n=4$ terms it is 1.07
and with $n=8$ terms it is 1.21.
At high temperatures, the high temperature expansions
are more sensitive to the limits used for the phase space integrals
than is the energy series form.


The dimer grand potential is typically about 5--10 times smaller
than the monomer grand potential at the same temperature.
For the loop grand potential,
the symmetrization factor induces an effective interaction
between the adjacent particles around the loop,
typically of the form $\cos ( Q_{j,j+1} P_{j,j+1})$
or $\sin ( Q_{j,j+1} P_{j,j+1})$.
Since rapidly oscillating terms tend to cancel,
it was found efficacious to introduce a cut-off $R_\mathrm{cut}^{p/q}$
and to neglect configurations
unless both $ |Q_{j,j+1}| \le R_\mathrm{cut}^{q}$
and $|P_{j,j+1}| \le R_\mathrm{cut}^{p}$.
For the present simple harmonic oscillator,
a value of $R_\mathrm{cut}^{p/q} = 4$ was found to change
the results by about 2\% at $\beta=0.1$
and by less than .1\% at $\beta =1$,
while substantially reducing the computation time.
(The results in the inset of Fig.~\ref{Fig:OmB}
do not use a cut-off.)

\subsection{Commutation Function and Averages}

The focus above has been on the grand potential,
and the commutation function was treated with that in mind.
In the case of the statistical average of an operator,
the specific commutation function can depend on the particular operator.
(The following analysis differs from that given in
Ref.~\onlinecite{Attard16}.)

The average of an operator is
\begin{eqnarray}
\left<  A \right>_{z,T}^\pm
& = &
\frac{1}{\Xi^\pm} \mbox{TR}'\, e^{-\beta \hat{\cal H}} \hat A
 \\ & = &
\frac{1}{\Xi^\pm}
\sum_{N=0}^\infty z^N
\sum_{\bf n}\!'
e^{-\beta E_{\bf n}}  A_{{\bf n}{\bf n}}
\nonumber \\ & = &
\frac{1}{\Xi^\pm}
\sum_{N=0}^\infty \frac{z^N}{N!}
\sum_{\bf n} \chi_{\bf n}^\pm
e^{-\beta E_{\bf n}} A_{{\bf n}{\bf n}}
\nonumber \\ & = &
\frac{1}{\Xi^\pm}
\sum_{N=0}^\infty \frac{z^N}{N!}
\sum_{\bf n}
\sum_{\hat{\mathrm P}} (\pm 1)^p
\langle {\bf n} | e^{-\beta \hat{\cal H}} \hat A
| {\hat{\mathrm P}{\bf n}}\rangle  .\nonumber
\end{eqnarray}
Suppose that the operator to be averaged is an ordinary function
of the position and momentum operators,
$\hat A = A(\hat{\bf p},\hat{\bf q})$.
The analysis proceeds as in the text, with
$ e^{-\beta \hat{\cal H}}  \Rightarrow  e^{-\beta \hat{\cal H}} \hat A$.
One ends up with
 \begin{eqnarray}
\left<  A \right>_{z,T}^\pm
& = &
\frac{1}{\Xi^\pm}
\sum_{N=0}^\infty \frac{z^N}{h^{dN}N!}
\sum_{\hat P} (\pm 1)^p
\int \mathrm{d}{\bf q} \, \mathrm{d}{\bf p}\;
 \\ && \mbox{ } \times
\frac{ \left\langle{\bf q} \left|
e^{-\beta \hat {\cal H} } \hat A
\right| {\bf p} \right\rangle
}{\left\langle {\bf q} | {\bf p} \right\rangle }
\frac{
\left. \left\langle \hat{\mathrm P} {\bf p} \right| {\bf q} \right\rangle \,
}{\left\langle {\bf p} | {\bf q} \right\rangle \,}
\nonumber \\ & \equiv &
\frac{1}{\Xi^\pm}
\sum_{N=0}^\infty \frac{z^N}{h^{dN}N!}
\int \mathrm{d}{\bf q} \, \mathrm{d}{\bf p}\;
\nonumber \\ && \mbox{ } \times
 e^{-\beta {\cal H}({\bf p},{\bf q})}\, A({\bf p},{\bf q})\,
 W_{A,p}({\bf p},{\bf q})\,
 \eta^\pm_q({\bf p},{\bf q}).\nonumber
\end{eqnarray}
Here the weight function for the average has been defined,
which, as in the text, can be written as
\begin{eqnarray}
\lefteqn{
e^{-\beta{\cal H}({\bf p},{\bf q})} A({\bf p},{\bf q}) W_{A,p}({\bf p},{\bf q})
} \nonumber \\
& = &
\frac{\langle{\bf q}|e^{-\beta\hat{\cal H}} \hat A |{\bf p}\rangle
}{\langle{\bf q}| {\bf p}\rangle }
\nonumber\\ & = &
\frac{1}{\langle{\bf q}| {\bf p}\rangle }
\sum_{\bf n}
\langle{\bf q}|e^{-\beta\hat{\cal H}} |{\bf n}\rangle \,
\langle{\bf n} |\hat A | {\bf p}\rangle
\nonumber\\ & = &
\frac{1}{\langle{\bf q}| {\bf p}\rangle }
\sum_{\bf n }e^{-\beta E_{\bf n}}
\langle {\bf q}| {\bf n}\rangle \,
\langle{\bf n} |\hat A | {\bf p}\rangle .
\end{eqnarray}
Conversely, since the original trace form for the statistical average
is unchanged by the cyclic permutation of the operators,
the statistical average can also be obtained from
\begin{eqnarray}
\lefteqn{
e^{-\beta{\cal H}({\bf p},{\bf q})} A({\bf p},{\bf q}) W_{A,q}({\bf p},{\bf q})
} \nonumber \\
& = &
\frac{\langle{\bf p}| e^{-\beta\hat{\cal H}} \hat A|{\bf q}\rangle
}{\langle{\bf p}| {\bf q}\rangle }
\nonumber\\ & = &
\frac{1}{\langle{\bf p}| {\bf q}\rangle }
\sum_{\bf n }e^{-\beta E_{\bf n}}
\langle {\bf p}|  {\bf n}\rangle \,
\langle{\bf n} |\hat A |{\bf q}\rangle.
\end{eqnarray}
With this the statistical average is equally well written
 \begin{eqnarray}
\left<  A \right>_{z,T}^\pm
& = &
\frac{1}{\Xi^\pm}
\sum_{N=0}^\infty \frac{z^N}{h^{dN}N!}
\int \mathrm{d}{\bf q} \, \mathrm{d}{\bf p}\;
 \\ && \mbox{ } \times
 e^{-\beta {\cal H}({\bf p},{\bf q})}\, A({\bf p},{\bf q})\,
 W_{A,q}({\bf p},{\bf q})\,
 \eta^\pm_p({\bf p},{\bf q}).\nonumber
\end{eqnarray}
One should \emph{not} assume that $W_{A,p} = W_{A,q}$.
If one swaps the order of the Maxwell-Boltzmann operator
and the operator to be averaged,
then these are replaced by $\tilde W_{A,p}$ and  $ \tilde W_{A,q}$,
with $\tilde W_{A,p} = W_{A,q}^* $ and $\tilde W_{A,q} = W_{A,p}^* $.

If the operator is only a function of position then the second form becomes
\begin{equation}
e^{-\beta{\cal H}({\bf p},{\bf q})} W_{A,q}({\bf p},{\bf q}) A({\bf q})
=
\frac{A({\bf q})}{\langle{\bf p}| {\bf q}\rangle }
\sum_{\bf n }e^{-\beta E_{\bf n}}
\langle {\bf p}| {\bf n}\rangle \,
\langle{\bf n} | {\bf q}\rangle .
\end{equation}
Hence if $\hat A = A({\bf \hat q})$, then
$ W_{A,q}({\bf p},{\bf q}) = W_q({\bf p},{\bf q})$.
If the operator is only a function of momentum, then the first form becomes
\begin{equation}
e^{-\beta{\cal H}({\bf p},{\bf q})} W_{A,p}({\bf p},{\bf q}) A({\bf p})
=
\frac{A({\bf p})}{\langle{\bf q}| {\bf p}\rangle }
\sum_{\bf n }e^{-\beta E_{\bf n}}
\langle {\bf q}| {\bf n}\rangle \,
\langle{\bf n} | {\bf p}\rangle .
\end{equation}
Hence if $\hat A = A({\bf \hat p})$,
then $ W_{A,p}({\bf p},{\bf q}) = W_p({\bf p},{\bf q})$.

These mean that in these two cases (or their linear combination)
one can use the original
commutation function $W_p({\bf \Gamma})$ or $W_q({\bf \Gamma})$,
and it would therefore be legitimate to regard it
as a weight function for classical phase space.

If the operator is a function of the energy operator
then one also has  a relatively simple form,
\begin{eqnarray}
\lefteqn{
e^{-\beta{\cal H}({\bf p},{\bf q})} W_{A,p}({\bf p},{\bf q})
A({\cal H}({\bf p},{\bf q}))
} \nonumber \\
& = &
\frac{1}{\langle{\bf q}| {\bf p}\rangle }
\sum_{\bf n } e^{-\beta E_{\bf n}} A(E_{\bf n})\,
\langle {\bf q}| {\bf n}\rangle \,
\langle{\bf n} | {\bf p}\rangle .
\end{eqnarray}
Hence if $\hat A = A (\hat{\cal H})$,
one sees that $\tilde W_{A,p} =  W_{A, p}$,
and that $W_{A,p}^* = W_{A, q} = \tilde W_{A, q}$.

This result is instructive in the linear case,
where  $A $ is the energy function itself,
$ \hat{\cal H} =  U(\hat {\bf q}) +  \hat p^2/2m$.
In this case
\begin{eqnarray}
\lefteqn{
e^{-\beta{\cal H}({\bf p},{\bf q})}
{\cal H}({\bf p},{\bf q})
W_{{\cal H},p}({\bf p},{\bf q})
} \nonumber \\
& = &
\frac{1}{\langle{\bf q}| {\bf p}\rangle }
\sum_{\bf n } e^{-\beta E_{\bf n}} E_{\bf n}\,
\langle {\bf q}| {\bf n}\rangle \,
\langle{\bf n} | {\bf p}\rangle .
\end{eqnarray}
But from the earlier analysis
one also has that
\begin{eqnarray}
e^{-\beta{\cal H}({\bf p},{\bf q})}
\lefteqn{
\left\{ W_{{\cal K},p}({\bf p},{\bf q}) {\cal K}({\bf p})
+ W_{{U},q}({\bf p},{\bf q}) U({\bf q})
\right\}
} \nonumber \\
& = &
\frac{{\cal H}({\bf p},{\bf q})}{\langle{\bf q}| {\bf p}\rangle }
\sum_{\bf n } e^{-\beta E_{\bf n}}
\langle {\bf q}| {\bf n}\rangle \,
\langle{\bf n} | {\bf p}\rangle
\nonumber \\ & = &
e^{-\beta{\cal H}({\bf p},{\bf q})}
{\cal H}({\bf p},{\bf q})
 W_{p}({\bf p},{\bf q}).
\end{eqnarray}
Note that $ W_{{\cal K},p} = W_p$ and $W_{U,q} = W_q$.
Hence the average energy can also be written
\begin{eqnarray} 
\lefteqn{
\left< \hat {\cal H} \right>_{z,\beta,V}^\pm
}  \\
& = &
\left< \hat {\cal K} \right>_{z,\beta,V}^\pm
+
\left< \hat U \right>_{z,\beta,V}^\pm
\nonumber \\ & = &
\frac{1}{\Xi^\pm}
\sum_{N=0}^\infty \frac{z^N}{h^{dN} N!}
\int \mathrm{d}{\bf \Gamma}\;
e^{-\beta{\cal H}({\bf \Gamma})}
\left\{ {\cal K}({\bf p}) W_p({\bf \Gamma})  \eta^\pm_q({\bf \Gamma})
\right. \nonumber \\ && \left. \mbox{ }
+ U({\bf q}) W_q({\bf \Gamma})  \eta^\pm_p({\bf \Gamma})
\right\}
\nonumber \\ & = &
\frac{1}{\Xi^\pm}\sum_{N=0}^\infty \frac{z^N}{h^{dN} N!}
\int \mathrm{d}{\bf \Gamma}\;
e^{-\beta{\cal H}({\bf \Gamma})}
 {\cal H}({\bf p}) W_p({\bf \Gamma})  \eta^\pm_q({\bf \Gamma}) . \nonumber
\end{eqnarray}
The final equality follows by taking the complex conjugate
of the potential energy average.
Alternatively using the complex conjugate of the average kinetic energy
corresponds to the replacement $W_p \eta^\pm_q \Rightarrow W_q \eta^\pm_p $.

This result means that the average energy can be equally taken
with either $W_{{\cal H},p}$ or with $W_p$
but it does not mean that $ W_{{\cal H},p} = W_p$.
In fact, one can confirm directly from the definitions that
$W_{{\cal H},p} = W_p - {\cal H}^{-1} \partial W_p/\partial \beta$.
One must therefore have that
$\sum_N (z^N/h^{dN}N! )\int \mathrm{d}{\bf \Gamma} \;
 e^{-\beta {\cal H}({\bf \Gamma})}\, \eta_q^\pm({\bf \Gamma})
\partial W_p({\bf \Gamma}) /\partial \beta = 0$.
In fact one can show that this holds individually
for each term $N$ and each permutation $\hat {\mathrm P}$).\cite{Attard18}
This is necessary for thermodynamic consistency,
since
the most likely energy is the temperature derivative of the grand potential,
$ \overline E = {\partial (\beta \Omega)}/{\partial \beta}$.
(See Ref.~\onlinecite{Attard18} for a more detailed treatment of averages.)


The above analysis reveals
that in the case of the average of an operator
that is a linear combination of `pure' functions
(a pure function depends
only on the position operator or only on the momentum operator),
then the original commutation function $W$ suffices to obtain
the statistical average.
This is rather useful from the computational viewpoint.



For a Hermitian operator $ A(\hat{\bf p},\hat{\bf q})$,
the quantum statistical average is real,
as is $A({\bf p},{\bf q})$.
Since  $\eta^\pm_x({\bf p},{\bf q})^* = \eta^\pm_x(-{\bf p},{\bf q})$,
it is sufficient that
$ A({\bf p},{\bf q})  W_{A,x}({\bf p},{\bf q})^*
= A(-{\bf p},{\bf q})  W_{A,x}(-{\bf p},{\bf q})$, $x=p,q$,
in order for the imaginary part to integrate to zero.

In any case,  the average must be real,
and so the imaginary part of the integrand must always integrate to zero
no matter how one formulates the integrand.
Nevertheless,
it may be more efficient computationally to write it
in the most symmetric fashion using the fact that the trace operation
is insensitive to the order of the operators
or of the position and momentum eigenfunctions.
Hence one can use the symmetric integrand
\begin{eqnarray}
\lefteqn{
 e^{-\beta {\cal H}({\bf \Gamma})}\,
 A({\bf \Gamma})
\overline W_A({\bf \Gamma})
\overline \eta^\pm({\bf \Gamma})
}   \\
&  \equiv &
\frac{1}{4} \left\{
\frac{  \left\langle  {\bf q} \left|
e^{-\beta \hat {\cal H} } \hat A
+
 \hat A  e^{-\beta \hat {\cal H} }
\right| {\bf p} \right\rangle
}{\langle  {\bf q} | {\bf p} \rangle }
\eta^\pm_q({\bf \Gamma})
\right. \nonumber \\  &  & \left. \mbox{ }
+
\frac{  \left\langle  {\bf p} \left|
e^{-\beta \hat {\cal H} } \hat A
+
 \hat A  e^{-\beta \hat {\cal H} }
\right| {\bf q} \right\rangle
}{\langle  {\bf p} | {\bf q} \rangle } \eta^\pm_p({\bf \Gamma})
\right\}
\nonumber \\ & = &
\frac{1}{2} \mbox{Re} \left\{
\frac{  \left\langle  {\bf q} \left|
e^{-\beta \hat {\cal H} } \hat A
+
 \hat A  e^{-\beta \hat {\cal H} }
\right| {\bf p} \right\rangle
}{\langle  {\bf q} | {\bf p} \rangle } \eta_q^\pm({\bf \Gamma})
\right\} \nonumber \\ & = &
\frac{1}{2}
 e^{-\beta {\cal H}({\bf \Gamma})}\, A({\bf \Gamma})
 \mbox{Re} \left\{
\left[ W_{A,p}({\bf \Gamma}) + \tilde W_{A,p}({\bf \Gamma}) \right]
\eta^\pm_q({\bf \Gamma})
\right\} .\nonumber
\end{eqnarray}
(See Ref.~\onlinecite{Attard18} for a more detailed treatment of averages.)


\subsubsection{Simple Harmonic Oscillator}

\begin{figure}[t!]
\centerline{
\resizebox{8cm}{!}{ \includegraphics*{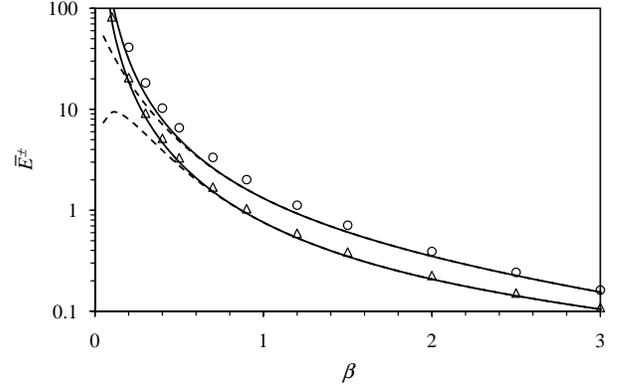} } }
\caption{\label{Fig:EvB}
Average energy in units of $\hbar \omega$ at $z=1$
for ideal simple harmonic oscillators in one dimension
using monomer and dimer loops (curves),
or else all loops up to 50-mers (symbols).
The upper data are for bosons and the lower data are for fermions.
The solid curves and the symbols are the exact analytic result,
Eq.~(\ref{Eq:olE-SHO}),
and the dashed curves  result from numerical quadrature over phase space,
Eq.~(\ref{Eq:<H>}),
using the commutation function with $n=8$ energy states,
and without a cut-off for the dimers.
}
\end{figure}

From thermodynamics, the most likely energy is given by
the temperature derivative of the grand potential,
$\overline E = {\partial (\beta \Omega)}/{\partial \beta}$.
Applying this to the exact analytic expression
for the simple harmonic oscillator,
Eq.~(\ref{Eq:BOmega-SHO}), yields
\begin{eqnarray}
\overline E^\pm & = & \label{Eq:olE-SHO}
\sum_{l=1}^\infty
(\pm 1)^{l-1} z^l  \frac{ d\hbar\omega }{2}
\left[ \frac{e^{-l\beta\hbar \omega/2}}{1-e^{-l\beta\hbar \omega}}
\right]^{d-1}
 \nonumber \\ & & \mbox{ } \times
\frac{  e^{-l\beta\hbar \omega/2} +  e^{-3l\beta\hbar \omega/2}
}{[1-e^{-l\beta\hbar \omega}]^2}
\nonumber \\ & \equiv &
\overline E_1 \pm \overline E_2 + \ldots
\end{eqnarray}

For the present  phase space formulation,
the individual terms in the loop expansion yield
\begin{eqnarray} \label{Eq:<H>}
\lefteqn{
\left<  {\cal H} \right>_{z,T;l}^\pm
} \nonumber \\
& = &
\frac{z^l }{h^{dl}  l}
\int \mathrm{d}{\bf \Gamma}^{l}
e^{-\beta {\cal H}^{(l)}({\bf \Gamma}^{l})}\,
\eta^{\pm(l)}({\bf \Gamma}^{l})\,
W^{(l)}({\bf \Gamma}^{l}) \,  {\cal H}^{(l)}({\bf \Gamma}^{l})
\nonumber \\ &= &
\frac{(\pm 1)^{l-1} z^l }{h^{dl}  l}
\int \mathrm{d}{\bf \Gamma}^{l}
\prod_{j=1}^l
\left[ e^{-\beta {\cal H}^{(1)}({\bf \Gamma}_j)}\,
 W^{(1)}({\bf \Gamma}_j)
\right.  \nonumber \\ & & \mbox{ } \times \left.
e^{ -({\bf q}_j-{\bf q}_{j+1}) \cdot {\bf p}_j /i\hbar }
\right]
\sum_{k=1}^l {\cal H}^{(1)}({\bf \Gamma}_k).
\end{eqnarray}
(These are either $W_p$ and $\eta^\pm_q$,
or else $W_q$ and $\eta^\pm_p$.)
These are for an ideal system.
The average can be given in terms of $W_{\cal H}$
by making the replacement in the integrand
\begin{equation}
\prod_{j=1}^l [ W^{(1)}_j ]
\sum_{k=1}^l {\cal H}^{(1)}_k
\Rightarrow
\sum_{k=1}^l W^{(1)}_{{\cal H},k} {\cal H}^{(1)}_k
\prod_{j=1}^l \! ^{(j\ne k)} W^{(1)}_j  .
\end{equation}
The above result has been confirmed by deriving it both
from the temperature derivative,
and from the original trace form for the statistical average
(not shown).

The average energy for the simple harmonic oscillator
is shown in Fig.~\ref{Fig:EvB}
as either the sum of the monomer and dimer terms
in these expressions,
or else the exact analytic expression (up to 50-mers).
It can be seen that bosons have a higher energy than fermions
at a given temperature.
Presumably Fermi exclusion gives rise to an effective repulsion
that leads to fewer fermions than bosons for a given fugacity.
It can be seen that at the present fugacity,  $z=1$,
dimers account for most of the difference between the two types of particles.
The separation between the two cases
would increase with increased fugacity,
as would the contribution from higher order loops.
It can be seen that
the phase space expression for the average energy is quite accurate.
The error at high temperatures could be ameliorated
by including more terms in the energy series
for the commutation function,
or else by using the high temperature expansions.
Imposing a cut-off of 4 changed the results by less than 0.03\%
at $\beta =0.2$.
The results using the commutation function $W$
were indistinguishable
(to at least six significant figures at all temperatures)
from those using $W_{\cal H}$,
which confirms the analysis earlier in this sub-section.

%
\section{Harmonic Approximation for Interacting Particles} \label{Sec:HarmInt}
\setcounter{equation}{0} \setcounter{subsubsection}{0}
%

One can use the above results to approximate
the commutation function of an interacting system
by casting each configuration as a collection of non-interacting
harmonic oscillators.
The interacting Hamiltonian consists
of the usual kinetic and potential energies,
\begin{equation}
{\cal H}({\bf p},{\bf q}) = {\cal K}({\bf p}) + U({\bf q}) ,
\end{equation}
with the latter containing many-body terms,
\begin{eqnarray}
U({\bf q}) & = &
\sum_{j=1}^N u^{(1)}({\bf q}_j)
+ \sum_{j<k}^N u^{(2)}({\bf q}_j,{\bf q}_k)
\nonumber \\ && \mbox{ }
+ \sum_{j<k<l}^N u^{(3)}({\bf q}_j,{\bf q}_k,{\bf q}_l)
+ \ldots
\end{eqnarray}
Distributing the energy equally,
the energy of particle $j$ 
can be defined as
\begin{eqnarray}
U_j({\bf q})
& = &
u^{(1)}({\bf q}_j)
+ \frac{1}{2} \sum_{k=1}^N\!^{(k\ne j)} \, u^{(2)}({\bf q}_j,{\bf q}_k)
\nonumber \\ && \mbox{ }
+ \frac{1}{3}
\sum_{k<l}^N\!^{(k,l\ne j)} \, u^{(3)}({\bf q}_j,{\bf q}_k,{\bf q}_l)
+ \ldots
\end{eqnarray}
The total potential energy is just
$ U({\bf q}) = \sum_{j=1}^N U_j({\bf q})$.

For a given configuration ${\bf q}$,
define the test energy for particle $j$,
$U_j({\bf r};{\bf q}) \equiv
\left. U_j({\bf q}) \right|_{{\bf q}_j={\bf r}}$.
Here the $j$th particle has been moved to ${\bf r}$,
all other particles remaining fixed in their positions
for the current configuration.
The location of the nearest
local minimum for particle $j$, $\overline {\bf q}_j({\bf q})$, satisfies
$ \left. \nabla_{\bf r} U_j({\bf r};{\bf q})
\right|_{{\bf r}=\overline {\bf q}_j}
={\bf 0}$.
Define the $d \times d$ second derivative matrix for particle $j$
at this local minimum as
\begin{equation}
\left. \nabla_{\bf r} \nabla_{\bf r} U_j({\bf r};{\bf q})
\right|_{{\bf q}_j=\overline {\bf q}_j}
\equiv
\overline{\underline{\underline U}}_j''.
\end{equation}
Newton's method for finding $\overline {\bf q}_j({\bf q})$
is discussed below.

For liquids and solids, a local minimum is well defined,
since each molecule is instantaneously caged by surrounding molecules.
Since the distance to the local minimum is likely to be small,
a second order expansion of the potential will be accurate.
As shown in Fig.~\ref{Fig:eW-Q},
the commutation function departs most from unity
close to the minimum in the potential.
For those unlikely configurations where a particle is not close
to a local minimum, the commutation function can be taken to be unity;
in such cases the Maxwell-Boltzmann factor will mean
these configurations have little weight in the overall average.

The potential energy of particle $j$ 
may be expanded to second order about the local minimum,
\begin{equation}
U_j({\bf q})
=
\overline U_j({\bf q})
+ \frac{1}{2} [{\bf q}_j-\overline{\bf q}_j][{\bf q}_j-\overline{\bf q}_j]
: \overline{\underline{\underline U}}_j'' ,
\end{equation}
where $\overline U_j({\bf q}) \equiv U_j(\overline{\bf q}_j;{\bf q})$.
The total potential energy is just the sum of these.

The second derivative  $d\times d$ matrix is positive definite
with eigenvalues $\lambda_{j\alpha}({\bf q})  > 0$,
and  orthonormal eigenvectors
$ \overline{\underline{\underline U}}_j'' {\bf X}_{j\alpha}
= \lambda_{j\alpha}{\bf X}_{j\alpha}$, $\alpha = x,y,\ldots, d$.
As $d$ is typically 1, 2, or 3, it is trivial numerically
to find these eigenvalues and eigenvectors.
The orthogonal matrix $\underline{\underline X}_j
= \{{\bf X}_{jx},{\bf X}_{jy},\ldots,{\bf X}_{jd}\}$
gives $\underline{\underline X}_j^\mathrm{T}
\overline{\underline{\underline U}}_j'' \underline{\underline X}_j
= \underline{\underline D}_j $,
where $\underline{\underline D}_j$ is a diagonal matrix.
For molecule $j$ in configuration ${\bf q}$ the eigenvalues
define the frequencies
\begin{equation}
\omega_{j\alpha}({\bf q}) = \sqrt{ \lambda_{j\alpha}({\bf q}) /m }
, \;\; \alpha = x, y, \ldots , d .
\end{equation}
With this the potential energy is
\begin{eqnarray}
U({\bf q})
& = &
\sum_{j=1}^N \overline U_j
+
 \frac{1}{2}\sum_{j=1}^N
({\bf q}_j-\overline{\bf q}_j) ({\bf q}_j-\overline{\bf q}_j) :
\overline{\underline{\underline U}}_j''
\nonumber \\ & = &
\sum_{j=1}^N  \overline U_j
+
\frac{1}{2} \sum_{j,\alpha}  \hbar \omega_{j\alpha} Q_{j\alpha}^2 .
\end{eqnarray}
Here $ Q_{j\alpha} \equiv
\sqrt{ { m \omega_{j\alpha} }/{\hbar} } \, Q_{j\alpha}' $,
and
$ {\bf Q}_j' =
 \underline{\underline X}_j^\mathrm{T}  [{\bf q}_j-\overline{\bf q}_j] $.
Also define
$\tilde q_{j\alpha} \equiv \sqrt{\hbar/m\omega_{j\alpha}} \, Q_{j\alpha}$.

With this harmonic approximation
for the potential energy,
the Hamiltonian in a particular configuration can be written
\begin{equation}
{\cal H}({\bf p},{\bf q})
=
\sum_{j=1}^N \overline U_j
+
\frac{1}{2} \sum_{j,\alpha}  \hbar \omega_{j\alpha}
\left[ P_{j\alpha}^2 + Q_{j\alpha}^2 \right] ,
\end{equation}
where  $ P_{j\alpha} = p_{j\alpha} /\sqrt{m\hbar\omega_{j\alpha}} $.
This represents $dN$ independent oscillators.
Hence the commutation function for the interacting system
for a particular configuration
can be approximated as the product of commutation functions
for effective non-interacting harmonic oscillators,
\begin{eqnarray}
\lefteqn{
e^{-\beta {\cal H}({\bf p},{\bf q})}
W_p({\bf p},{\bf q})
}  \\
& \approx &
\prod_{k=1}^N
e^{-\beta \overline U_k }
\prod_{j,\alpha}
 e^{- \beta\hbar \omega_{j\alpha}[P_{j\alpha}^2+Q_{j\alpha}^2]/2 }
W_{p;j\alpha}({\bf p},{\bf q}) ,\nonumber
\end{eqnarray}
with the single particle, one-dimensional commutation function
being given by
\begin{eqnarray}
\lefteqn{
e^{- \beta\hbar \omega_{j\alpha}[P_{j\alpha}^2+Q_{j\alpha}^2]/2 }
W_{p;j\alpha}({\bf p},{\bf q})
}  \\
& = &
e^{q_{j\alpha}p_{j\alpha}/i\hbar}
\sum_{n_{j\alpha}=0}^\infty
e^{- \beta \hbar \omega_{j\alpha} [ n_{j\alpha} + .5] }
 \check \phi_{n_{j\alpha}}(p_{j\alpha})
\phi_{n_{j\alpha}}(\tilde q_{j\alpha})
 \nonumber \\ & = &
e^{q_{j\alpha}p_{j\alpha}/i\hbar}
e^{-P_{j\alpha}^2/2}
e^{-Q_{j\alpha}^2/2}
e^{- \beta \hbar \omega_{j\alpha} /2 }
\nonumber \\ && \mbox{ } \times
\sqrt{2}
\sum_{n_{j\alpha}=0}^\infty
\frac{i^{n_{j\alpha}}
e^{- \beta \hbar \omega_{j\alpha}  n_{j\alpha} }
}{
2^{n_{j\alpha}} {n_{j\alpha}}!  }
\mathrm{H}_{n_{j\alpha}}(P_{j\alpha})
\mathrm{H}_{n_{j\alpha}}(Q_{j\alpha}) .\nonumber
\end{eqnarray}
The imaginary terms here are odd in momentum.
Note that the prefactor $ e^{q_{j\alpha}p_{j\alpha}/i\hbar}$
comes from the factor $ e^{{\bf q} \cdot {\bf p}/i\hbar}$
from the phase space transformation
factor $1/\langle {\bf q}|{\bf p} \rangle$,
which is independent of the simple harmonic oscillator approximation.
Based on the numerical results presented in the preceding section,
one only needs to keep a few terms in this series
and to evaluate it for particles
that are close to their local potential minimum.


\subsubsection{Newton's Method}

One can approximate the location of the local minimum for particle $j$
in configuration ${\bf q}$ as follows.
Take the second derivative matrix to be
\begin{equation}
\overline{\underline{\underline U}}_j''
\approx
{\underline{\underline U}}_j''
\equiv
\nabla_j \nabla_j U_j({\bf q}) .
\end{equation}
Then with
\begin{eqnarray}
U_j({\bf q}) & = &
U_j({\bf r})|_{\overline{\bf q}_j}
+ [{\bf q}_j-\overline{\bf q}_j] \cdot
\nabla U_j({\bf r})|_{\overline{\bf q}_j}
 \\ && \mbox{ }
+ \frac{1}{2} [{\bf q}_j-\overline{\bf q}_j][{\bf q}_j-\overline{\bf q}_j]
: \nabla \nabla U_j({\bf r})|_{\overline{\bf q}_j} , \nonumber
\end{eqnarray}
the derivative is,
\begin{eqnarray}
\nabla_j U_j({\bf q}) & = &
\nabla U_j({\bf r})|_{\overline{\bf q}_j}
+  \nabla \nabla U_j({\bf r})|_{\overline{\bf q}_j}
\cdot [{\bf q}_j-\overline{\bf q}_j]
\nonumber \\ & \approx &
\nabla U_j(\overline{\bf q}_j)
+ {\underline{\underline U}}_j'' [{\bf q}_j-\overline{\bf q}_j].
\end{eqnarray}
Setting $\nabla U_j(\overline{\bf q}_j) = {\bf 0}$ yields
\begin{equation}
\overline{\bf q}_j
\approx
{\bf q}_j-({\underline{\underline U}}_j'')^{-1} \nabla_j U_j({\bf q})  .
\end{equation}
One can refine this estimate by successive approximation,
\begin{equation}
\overline{\bf q}_j^{(n+1)}
\approx
\overline{\bf q}_j^{(n)}
-\left[{\underline{\underline U}}_j''(\overline{\bf q}_j^{(n)})\right]^{-1}
\nabla_j U_j(\overline{\bf q}_j^{(n)})  .
\end{equation}

%
\section{Conclusion}
\setcounter{equation}{0} \setcounter{subsubsection}{0}
%

In this paper a formally exact transformation has been given
that expresses quantum statistical mechanics
as an integral over classical phase space.
Two phase functions that reflect specific quantum effects
result:
a commutation function
that accounts for the non-commutativity
of position and momentum operators,
and a symmetrization function that accounts
for wave function symmetrization (bosons)
or  anti-symmetrization (fermions).
The latter is more computationally tractable
than conventional methods of treating occupation states,
such as Slater determinants.
To leading order (high temperatures, low densities),
the phase functions are unity
and the theory reduces exactly
to classical statistical mechanics.
The magnitude of the quantum effects
can be estimated by truncating the respective series expansions
for the commutation and symmetrization functions,
which provides a systematic and quantifiable
way to approximate quantum systems.

The present phase space method was illustrated and tested
for non-interacting quantum harmonic oscillators,
for which system exact analytic results are known.
It was demonstrated that the quantum grand potential
and the average quantum energy
could be obtained accurately from an integral over classical phase space.
Both the high temperature expansion and the energy series
for the commutation function were tested
and found to have overlapping regimes of reliability.
Surprisingly few terms were required to get accurate results
with the energy series at low temperatures.
It was also demonstrated that the dimer term in the symmetrization function
sufficed for bosons and for fermions
at intermediate and low temperatures
at a fugacity of $z=1$.

Numerical results for the commutation function
for the quantum harmonic oscillator
showed that it only departed from unity near a minimum in the potential.
This suggests that a mean field theory based
upon an expansion about local potential minima
together with the analytic oscillator results
should be accurate for quantum condensed matter, interacting particle systems.
It remains to explore the computational feasibility
of such a mean field approach
and to test it against, for example, the high temperature expansion
for the commutation function
that has previously been applied to interacting Lennard-Jones liquids.
\cite{STD2,Attard17,Attard16}

\subsubsection*{Notes Added}

(1.) An improved generic treatment of the factorization
of the symmetrization function for averages,
and a demonstration of the internal consistency of the approach,
are given in Ref.~\onlinecite{Attard18}.

(2.) Because of the compact nature of contributing permutation loops,
in general the symmetrization function only has to be calculated for
permutation loops consisting of consecutive nearest neighbors,
whence appropriate neighbor tables should further ameliorate
the computational burden.


\end{document}